# Multiuser Modulation Classification Based on Cumulants in AWGN Channel

Masoud Zaerin, Babak Seyfe, *Member, IEEE,* and Hamid R. Nikoofar

*Abstract*— In this paper the negative impacts of interference transmitters on automatic modulation classification (AMC) have been discussed. We proposed two approaches for AMC in the presence of interference: single user modulation classification (SUMC) and multiuser modulation classification (MUMC). When the received power of one transmitter is larger than the other transmitters, SUMC approach recognizes the modulation type of that transmitter and other transmitters are treated as interferences. Alternatively when the received powers of all transmitters are close to each other we propose MUMC method to recognize the modulation type of all of the transmitted signals. The features being used to recognize the modulation types of transmitters for both approaches, SUMC and MUMC are higher order cumulants. The superposition property of cumulants for independent random variables is utilized for SUMC and MUMC. We investigated the robustness of our classifier with respect to different powers of the received signals via analytical and simulation results and we have shown the analytical results will be confirmed by simulations. Also we studied the effect of signal synchronization error via simulation results in the both condition for MUMC and SUMC.

*Index Terms*— Automatic modulation classification (AMC), higher-order cumulant, interference, single user modulation classification (SUMC), multiuser modulation classification (MUMC).

I. INTRODUCTION

AUTOMATIC modulation classification (AMC) is a technique to identify modulation type of the received signal with a little prior knowledge about the parameters of signal such as carrier phase and frequency, symbol period, etc [1]-[2]. It plays a key role in various applications such as network traffic administration [1]-[2], intelligent modems [3]-[4], software defined radio [4]-[6], frequency spectrum monitoring, electronic surveillance [2], [5], electronic warfare and threat analysis [2], [4]-[5]. In a non-cooperative environment AMC is a difficult task, as no prior

Manuscript received August 14, 2009. This work was supported by Iran Telecommunication Research Center, Tehran, Iran under Grant T/500/7224.
M. Zaerin and B. Seyfe are with the Department of Electrical Engineering Shahed University, Tehran, Iran (e-mails: zaerin@shahed.ac.ir, seyfe@shahed.ac.ir). Corresponding author is B. Seyfe (e-mail: seyfe@shahed.ac.ir).
H. R. Nikoofar is with the Department of Electrical and Computer Engineering, Tarbiat Modares University, Tehran, Iran. (e-mail: hr.nikoofar@mci.ir).

knowledge of the incoming signal is available. For the classification algorithm in the AMC system there are two methods: the decision theoretic method that is based on the maximum likelihood function (ML) and the statistical pattern recognition (PR) methods. Based on chosen classification algorithm some preprocessing may be needed. This preprocessing may involve some of tasks like: carrier phase and frequency estimation, equalization, symbol period estimation, and noise reduction, etc. ML methods suffer from very high computational complexity [3]. These methods are optimal with Bayesian criterion and minimize the probability of error but, ML methods are not robust with respect to model mismatches such as phase and frequency offsets, residual channel effects, synchronization errors and deviation from noise distribution. In the statistical pattern recognition methods, the classifier at first extracts appropriate features from the received waveform and then uses these features to decision making and recognize the modulation type of the received signal. The method that based on pattern recognition may not be optimal but it is usually simple to implement and has low computational complexity. If it was designed properly, its performance can be suboptimal [4]. Some examples of features that used for AMC will be reviewed in the sequel. The variances of the amplitude, phase and frequency of the centered normalized signal [7], zero-crossing intervals [8], and magnitude of the signal wavelet transform (WT) after peak removal [9] are some examples about the variance measure. Also, the phase probability density function, i.e. PDF, [10] and its statistical moments have been used in [11]-[12]. The moments, cumulants, and cyclic cumulants of the observed signal were used as robust features against uncertainties in [1], [3]-[5], [13], [20], [22]-[25], [27], [28]. When the training data are available fuzzy logic [14] have been proposed. The entropy [15], a moment matrix technique [16] and a constellation shape recovery method [17] were also used as feature for AMC. Different methods were employed for decision making, such as PDF-based [3], [11]-[12], the Hellinger distance [18]-[19], the Euclidian distance [1]-[4], [20] and unsupervised clustering techniques [21]. A hierarchical framework based on fourth-order cumulants is proposed in [3]. Marchand et al. [22]

proposed a combination of fourth and second-order cyclic cumulant (CC) magnitudes for QPSK and QAM signals classification. The behaviors of cyclic cumulants with order up to four [23] and six [24] as features are investigated. In [20] Eighth-order cyclic cumulant are used to recognize QAM, PSK and ASK signals. The features based on CCs which robust to carrier frequency offset and phase noise were used to classify QAM signals in [4]. A multi-antenna CCs based classifier for digital linear modulation in flat fading channel was proposed in [1]. In [5], an AMC using forth order cumulant features in the multipath fading channel was proposed. In [25] the Cramer-Rao lower bound is derived for the fourth-order cumulant estimator when the modulation type candidates are BPSK and QPSK over the additive white Gaussian noise (AWGN) channel. Many existing techniques for digital modulation recognition are reviewed in [26] and it provided useful guideline for choosing appropriate classification algorithms for different modulations, from the large pool of available techniques. In [27] the authors proposed classification of MQAM/MPSK signals in multipath fading environments. The proposed approach, in which the two-step equalization strategy and higher-order cumulants based classifier are adopted to classify the MPSK and higher-order QAM signals. The authors in [28] propose to use the fourth order cumulants to distinguish OFDM from single carrier signals.

It is obvious that the interference results in the severe AMC performance degradation or induces large classification errors. In the most of existing researches, only the effect of one transmitter has been taken into account at the receiver and the effect of interference transmitters have been ignored [1]-[8], but in many applications such as multiple access channels or channels with interference this assumption is not satisfied. However in this paper, it is assumed that there are other transmitters that their signals are received at the receiver and therefore we have a multiuser channel. In this situation we have two approaches for AMC: *Single User Modulation Classification* (SUMC) and *Multiuser Modulation Classification* (MUMC). If the received power of one transmitter is larger than other transmitters, we used SUMC approach to recognize the modulation type of that transmitter and the others are treated as interferences. Alternatively in another condition where the received powers of all transmitters are close to each other (as a worst case condition) we propose MUMC method to recognize the modulation type of all of the present signals at the receiver. In this paper we used a cumulant-based method to perform SUMC and MUMC. These statistics characterize the shape of the distribution of the noisy baseband in phase and quadrature samples and we utilize their superposition property for independent random variables, in both SUMC and MUMC. The cumulants benefit the robustness against frequency offset, channel phase and phase jitter and additive white and colored Gaussian noise [1], [3], [4].

This paper is organized as follows. In section II we describe the model of the observed signal. In section III we describe single user modulation classification. We explain MUMC method in the worst case that all of transmitters have the same received powers at receiver, in section IV. Sensitivity of the proposed MUMC classifier with respect to the different power of received signals will be described in section V via analytical operations and simulations. We investigate the effect of synchronization error on the modulation classifications performance, via simulation results, in section VI and conclusions are drown in section VII.

## II. SIGNAL MODEL

Assume a multiuser channel in the presence of modulated signals and additive white Gaussian noise. The block diagram of such a system is shown in Fig. 1. At this scenario we have $M$ transmitters that each transmitter has one of the $Q$ candidated modulation types and each transmitter has the separated modulation type from others, therefore $Q \geq M$. Thus the received signal is given by

$$y(n) = \sum_{k=1}^{M} \sum_{\ell=-\infty}^{\infty} \alpha_k e^{j2\pi f_k nT + j\theta_k(n) + j\varphi_k} x_k(\ell) h_k(nT - \ell T + \varepsilon_{T_k} T) + g(n). \qquad (1)$$

Also $x_k(\ell)$ is the $k$-th transmitter (complex) symbol at $\ell$-th time interval, $h_k(.)$ represents the residual channel effects ( e.g., due to synchronization error), $\theta_k(n)$ is the phase jitter, $f_k$ is the frequency offset, $\alpha_k$ is an (unknown) amplitude factor of the $k$-th received signal and $\varphi_k$ is the channel phase of $k$-th transmitter, respectively. $T$ is the symbol period, $\varepsilon_{T_k}$ $(0 \leq \varepsilon_{T_k} < 1)$ represents synchronization error factors of $k$-th transmitter, $\{y(n)\}_{n=1}^{N}$ is the received waveform where $N$ is the number of observed symbols, and $g(n)$ is AWGN with zero mean and variance $\sigma^2$. The number of symbol points at the $i$-th constellation type is $N_i$ where $i = 1, \dots, Q$. We assume that each channel is independent of the others. Also the symbols in each transmitter are independent and identically distributed (i.i.d). Without any loss of generality we assume that all the transmitters have the same symbol rate and carrier frequency. Note that when multiple interfering signals are presented at the receiver, which overlap in time, frequency and space, but with different symbol rates, they can be separated based on the selectivity of cyclic cumulants (different cycle frequencies) [1]. Also we can separate the signals with different frequencies via a filter bank. If the carrier frequencies of the transmitters are not the same but close enough to each other that could not be separated via filter bank, it will be modeled by the frequency offset in the baseband. Also in general, we assume that the modulation constellations can be normalized and have unit variance, thus by assuming equiprobable and zero mean symbols we have, $E[|x_i(n)|^2] = N_i^{-1} \sum_{m=1}^{N_i} |s_i(m)|^2$ that equals to one for $i = 1, \dots, Q$, where $s_i(m)$ is the $m$-th alphabet at $i$-th modulation constellation, $E[.]$ is the expectation and $|.|$ denotes the absolute function [1].

According to the robustness of cumulants with respect to frequency offset, phase jitter and channel phase [1], [3] and [4], we ignore their effects and therefore (1) can be rewritten as

$$y(n) = \sum_{k=1}^{M} \sum_{\ell=-\infty}^{\infty} \alpha_k\, x_k(\ell)\, h_k\big(nT - \ell T + \varepsilon_{T_k} T\big) + g(n). \qquad (2)$$

Then in future we will use (2) as our signal model. In the next section we will study the SUMC performance.

## III. SINGLE USER MODULATION CLASSIFICATION

At this section we recognize the modulation type of the most received power signal at the receiver based on (2) and others are treated as interference. According to the negative effect of interferences transmitters we need to improve the classifier performance by features that compensate the effect of the interference. We use the combination of forth-order cumulant and sixth-order cumulant as feature to classify the modulation type.

### A. Definition of the features

For a complex-valued zero-mean stationary random process $y(n)$ the second-order\one-conjugate, forth-order\two-conjugate and the sixth-order\three-conjugate cumulants can be defined as followings, respectively, [1], [3]

$$C_{21}(y) = cum(y(n), y^*(n))$$
$$= E[|y(n)|^2] - |E[y(n)]|^2 \tag{3}$$

$$C_{42}(y) = cum(y(n), y(n), y^*(n), y^*(n))$$
$$= E[|y(n)|^4] - |E[y(n)^2]|^2 - 2(E[|y(n)|^2])^2 \tag{4}$$

$$C_{63}(y) = cum(y(n), y(n), y(n), y^*(n), y^*(n), y^*(n))$$
$$= E[|y(n)|^6] - 9E[|y(n)|^4]E[|y(n)|^2] - 3E[y^*(n)^3 y(n)]E[y(n)^2])$$
$$-3E[y^*(n)y(n)^3]E[y^*(n)^2] + 18E[y^*(n)^2]E[y(n)^2]E[|y(n)|^2]$$
$$+12(E[|y(n)|^2])^3, \tag{5}$$

where $*$ and $cum(.)$ denote the conjugation sign and cumulant operators, respectively. Due to investigate the negative effects of interference transmitters on the performance of our classifier, in this section we ignore the effects of synchronization error and we consider its effect in section VI. We use the composition of $C_{42}$ and $C_{63}$ as useful feature for modulation classification as follows

$$f_C = \left| \frac{C_{42}(y)}{(C_{63}(y))^{\frac{2}{3}}} \right|. \tag{6}$$

Since the symbol sequences of transmitters are independent from each other and noise, based on the model presented at (2), we can write (6) as [3]

$$f_C = \left| \frac{\sum_{k=1}^{M} C_{42}(\alpha_k\, x_k(n)) + C_{42}(g)}{\left(\sum_{k=1}^{M} C_{63}(\alpha_k\, x_k(n)) + C_{63}(g)\right)^{\frac{2}{3}}} \right|. \qquad (7)$$

According to (4), (5) due to the symmetric conjugation of $C_{42}$ and $C_{63}$ we have

$$C_{42}(\alpha_k\, x_k(n)) = |\alpha_k|^4 C_{42}(x_k(n)), \qquad (8)$$

Therefore according to (8), (9) we can write (7) as follow

$$f_C = \left| \frac{\sum_{k=1}^{M} |\alpha_k|^4 C_{42}(x_k(n))}{\left(\sum_{k=1}^{M} |\alpha_k|^6 C_{63}(x_k(n))\right)^{\frac{2}{3}}} \right|. \qquad (10)$$

Note that for additive Gaussian noise $C_{42}(g)$ and $C_{63}(g)$ are equal to zero [3]. By assuming that the modulation constellations have unit variance the power of each received signal is $|\alpha_k|^2, k = 1, 2, \ldots, M$. Let subscription $d$ denotes the desired transmitter, if its power is more than other transmitter sufficiently, we can ignore the effect of interference transmitters, $\sum_{\substack{1 \leq k \leq M \\ k \neq d}} |\alpha_k|^4 C_{42}(x_k(n))$ with respect to the effect of $|\alpha_d|^4 C_{42}(x_d(n))$ and also ignore the effect of $\sum_{\substack{1 \leq k \leq M \\ k \neq d}} |\alpha_k|^6 C_{63}(x_k(n))$ respect to the effect of $|\alpha_d|^6 C_{63}(x_d(n))$. Therefore we can write (10) as

In [3] a forth order cumulant based classifier was proposed and its properties are studied comprehensively as

$$f = \left|\frac{C_{42}(y)}{(C_{21}(y) - \sigma^2)^2}\right|. \quad (12)$$

We will show that in the presence of interference transmitters, our proposed method based on (11), with a marginal increase on the complexity computation can increase the performance of classifier in contrast to the reported method in [3] significantly. Moreover, based on (11) and (12), in the proposed feature there is no need to estimate the noise variance $\sigma^2$, in compared with the feature that used in [3].

B. *Feature estimation and decision making*

The cumulants in (4) and (5) can be estimated from the symbol estimates of the corresponding moments [3], [5]. We estimate our feature as

$$\hat{f}_c = \left|\frac{\hat{C}_{42}(y)}{\left(\hat{C}_{63}(y)\right)^{\frac{2}{3}}}\right|. \quad (13)$$

For decision-making we compare the estimated feature by (13), with the theoretical values of all candidate modulation features and then decide about the modulation type of the received signal. We have shown these theoretical values of features for each candidate modulation type in Table I. These values are computed with (4), (5) and (13), for the ideal noise free constellation and in the absence of interferences. In this table we assume that the modulation constellation symbols are equiprobable and have unit variance. For recognizing the modulation type of the received signal we compare the estimated feature via (13), with its theoretical value as follows

$$\hat{\imath} = arg \min_{i=1,\ldots,Q} |f_{ci} - \hat{f}_c| \quad (14)$$

where $i = 1, \ldots, Q$, is the candidate modulation type, $\hat{\imath}$ represents the estimated modulation type of the received signal, $f_{ci}$ is the theoretical feature of $i$-th modulation type.

## C. Performance of single user modulation classification

In this subsection the results of different simulations are presented to illustrate the performance of SUMC and compare our results with the results of [3] in the same condition. To define the classifier performance, we use the average probability of correct classification $p_{cc}$ as

$$p_{cc} = \sum_{i=1}^{Q} p_i\, p(i|i), \tag{15}$$

where $p_i$ is the probability that the modulation type of the desired transmitter is $i$ that means the other transmitters have lower received power and they are treated as interference. By assuming that a priori probability of modulation type of the desired transmitter for all modulation types candidate are the same, $p_i$ equals to $\frac{1}{Q}$. Let $p(\hat{\imath}|i)$ denotes the conditional probability when $i$ is the actual modulation type of the desired transmitter and we classify it as $\hat{\imath}$. It means that $p(i|i)$ is the probability of correct classification where $i$-th transmitter is the desired. The simulation results are based on signal to noise ratio (SNR) and signal to interference ratio (SIR) that are defined respectively as

$$SNR = 10\, \log_{10} \frac{var(\alpha_d\, x_d(n))}{\sigma^2}, \tag{16}$$

$$SIR = 10\, \log_{10} \frac{var(\alpha_d\, x_d(n))}{\sum_{\substack{1 \leq k \leq M \\ k \neq d}} var(\alpha_k\, x_k(n))}, \tag{17}$$

where $var(.)$ denotes the variance.

Simulation results are based on Monte Carlo method. For each Monte Carlo trial, the appropriate feature estimated based on $N$ data symbols. We assume four transmitters with different modulation types such as $BPSK$, $QPSK$, $4-PAM$ and $16-QAM$. It means that the number of the transmitters equal to the number of candidate modulation types. All results are based on 2000 Monte Carlo trials, i.e., 8000 trials for four classes. Fig. 2 compares the performances of SUMC and forth-order cumulant based classifier [3]. In this figure the probability of correct classification is illustrated versus $SNR$ with $SIR = 5, 10\ dB$ and $N = 2000$. It shows that our proposed method has better performance rather than forth-order cumulant based classifier. It is obvious that the performance of forth-order cumulant based classifier even in $SIR = 10$ dB is less than our proposed method in $SIR = 5$ dB (when SNR is higher than $-2.5$ dB).

## IV. MULTIUSER MODULATION CLASSIFICATION

In this section we want to recognize the modulation type of all transmitters by using cumulants in a multiuser channel. As we described in section II, at this scenario we have $M$ transmitters that each transmitter has one of the $Q$ candidate modulation types and each transmitter has the separated modulation type from others. Each set of $M$ possible transmitter's modulations called a *supper class*. The purpose of our classifier is the recognition of the modulation types of all the received signals by recognizing their super class. The total number of super classes with $M$ separated modulation types that collected from $Q$ candidate modulation types, can enumerates as

$$H = \binom{Q}{M} \tag{18}$$

$$= \frac{Q!}{M!\,(Q-M)!},$$

where $(.)!$ is the factorial operand. Then we have $H$ super class that each super class has $M$ modulation types. We illustrate $i$-th super class with $Sc_i = \{C_{i_1}, \ldots, C_{i_M}\}, i = 1, \ldots, H$. In this section we ignore the effect of synchronization error and we will consider its effect in section VI. According to (3) if we compute the second-order\one-conjugate cumulant of $y(n)$ we have

$$C_{21}(y) = \sum_{k=1}^{M} |\alpha_k|^2 C_{21}(x_k(n)) + \sigma^2 \tag{19}$$

According to (3), $C_{21}(x_k(n))$ is equal to $var(x_k(n))$, then by unit variance constellation assumption, (19) can be written as

$$C_{21}(y) = \sum_{k=1}^{M} |\alpha_k|^2 + \sigma^2 \tag{20}$$

We consider the worst case assumption that the powers of all the received signals are the same. In section V we will discuss about the sensitivity of the classifier when this assumption is not valid, analytically and via simulations, then (20) can be written as

$$C_{21}(y) = M |\alpha_k|^2 + \sigma^2, \quad k = 1, \dots, M \tag{21}$$

Hence

$$|\alpha|^2 \triangleq |\alpha_k|^2$$

$$= \frac{C_{21}(y) - \sigma^2}{M}, \quad k = 1, \dots, M \tag{22}$$

According to (22), (7) and (8) we have

$$C_{42}(y) = \sum_{k=1}^{M} |\alpha_k|^4 C_{42}(x_k(n))$$

$$= |\alpha|^4 \sum_{k=1}^{M} C_{42}(x_k(n)). \tag{23}$$

By combination of (22) and (23) it is possible to omit the effect of unknown amplitude, i.e. $|\alpha|^4$, as

$$f_{SC} = \sum_{k=1}^{M} C_{42}(x_k(n))$$

$$= \frac{C_{42}(y)}{\left(\frac{C_{21}(y) - \sigma^2}{M}\right)^2}, \tag{24}$$

where $f_{SC}$ is the super class feature. According to Table I as the forth-order cumulant of each modulation type is unique, the summation of cumulants of several modulation types is unique too and it can be used as a feature to recognize the modulations types in each super class.

To illustrate the performance of this method, the results of different simulations are presented. We assume that the symbol period, carrier frequency, pulse shape, noise variance and number of transmitters are known. We assume there are three transmitters with different modulation types that can be selected form $\{16-QAM, BPSK, QPSK, 4-PAM\}$. Therefore $M = 3$, $Q = 4$ and we have four super class such as $Sc_1 = \{BPSK, QPSK, 4-PAM\}$, $Sc_2 = \{16-QAM, QPSK, 4-PAM\}$, $Sc_3 = \{BPSK, 16-QAM, 4-PAM\}$ and $Sc_4 = \{BPSK, QPSK, 16-QAM\}$. The cumulants in (24) can be estimated from the estimates of the corresponding moments [3], [5]. Therefore super class feature can be estimated as

$$\hat{f}_{SC} = \frac{\hat{C}_{42}(y)}{\left(\frac{\hat{C}_{21}(y) - \sigma^2}{M}\right)^2}. \tag{25}$$

For recognizing the super class and therefore recognition the modulations types of the received signals, we compare the estimated feature via (25), with its theoretical value as follows

$$\hat{\imath} = arg \min_{i=1,\ldots,H} |f_{SC_i} - \hat{f}_{SC}|, \tag{26}$$

where $\hat{\imath}$ represents the estimated super class of the received signals, $f_{SC_i}$ is the theoretical feature of $i$-th super class $Sc_i$. According to (24) the theoretical value of feature for each super class can be computed by summation of the theoretical cumulants value of the modulation types that exist in corresponding super class. Therefore we have $f_{SC_1} = -4.36$, $f_{SC_2} = -3.04$, $f_{SC_3} = -4.04$, $f_{SC_4} = -3.68$. To define the multiuser classifier performance, we use the average probability of correct classification for supper class, $p_{Scc}$ as

$$p_{Scc} = \sum_{i=1}^{H} p_{Si} \, p_S(i|i), \tag{27}$$

where $p_{Si}$ is the probability that $i$-th super class occurs. Let all the super classes occur with the same probability then $p_{Si}$ equals to $\frac{1}{H}$. Let $p_S(\hat{\imath}|i)$ denotes the conditional probability when the actual super class $i$ is occurred and it is classified as $\hat{\imath}$, thus $p_S(i|i)$ is the probability of correct classification for $i$-th supper class. As it is desired to recognize the modulation type of all of the transmitted signals in each super class we define $SNR$ as

$$SNR = 10\ \log_{10} \frac{\sum_{k=1}^{M} var(\alpha_k\ x_k(n))}{\sigma^2}. \tag{28}$$

In fact the total power of all of the received signals, i.e., super class power, is evaluated in (28). All results are based on 2000 Monte Carlo trials, i.e., 8000 trials for four super class problems. Fig. 3 shows the probability of correct classification of super classes versus $SNR$ that parameterized by $N$. This figure illustrates the reasonable performance of multiuser classifier for various values of $N$. Note that in Fig. 3, due to equal unknown amplitude assumption that causes to equal received powers, $SIR$ is very low and equals to about $-3$ dB for each received signal.

V. SENSITIVITY OF MUMC APPROACH WITH RESPECT TO THE RECEIVED SIGNAL POWERS

At the above proposed MUMC we assumed the worst case model where all the received signals have the same power. In this section we consider the sensitivity of the proposed method with respect to the case where the received powers of the signals are not equal. In this case, we can write $f_{SC}$ by combining (20), (23), (24) as

$$f_{SC} = \frac{\sum_{k=1}^{M}|\alpha_k|^4 C_{42}(\ x_k(n))}{\left(\frac{\sum_{k=1}^{M}|\alpha_k|^2}{M}\right)^2}. \tag{29}$$

We will compute the sensitivity of this measure when the coefficients $\alpha_k, k = 1, 2, ..., M$ are not equal. This will be done via two examples analytically.

A. *Example 1: Sensitivity for super classes with two modulation types from three candidates*

We assume that the number of transmitted signals is $M = 2$ and $\alpha_2 = \alpha_1 + \epsilon_1$ where $\epsilon_1$ is the difference of the unknown amplitudes of these two received signals, therefore we can write (29) as

$$f_{SC} = \frac{|\alpha_1|^4 C_{42}(\ x_1(n)) + |\alpha_1 + \epsilon_1|^4 C_{42}(\ x_2(n))}{\left(\frac{|\alpha_1|^2 + |\alpha_1 + \epsilon_1|^2}{2}\right)^2}$$

$$= \frac{C_{42}(x_1(n)) + |1+\delta_1|^4 C_{42}(x_2(n))}{\left(\frac{1+|1+\delta_1|^2}{2}\right)^2}, \tag{30}$$

where $\delta_1 = \frac{\epsilon_1}{\alpha_1}$. Using (30) helps to find the undesired region for $\delta_1$, i.e. the error region. This means that if $\delta_1$ was placed in the error region, $f_{SC}$ will deviate from the correct decision boundary. We assume two transmitters with different modulation types can be selected form $\{BPSK, QPSK, 4-PAM\}$. We have three super class such as $Sc_1 = \{BPSK, QPSK\}$, $Sc_2 = \{QPSK, 4-PAM\}$ and $Sc_3 = \{BPSK, 4-PAM\}$ then according to (24) the theoretical features for these super classes are $f_{SC_1} = -3, f_{SC_2} = -2.36, f_{SC_3} = -3.36$, respectively. By substituting the theoretical values of cumulants of each supper class from the Table I, in (30) we have $f_{SC}$ as a function of $\delta_1$. For $\delta_1 \neq 0$ the values of $f_{SC}$ deviates from the theoretical values. To find the error region, according to (26), the decision boundary can be found when the absolute value of the features difference is equal to the mean of their theoretical features, i.e., decision point. Set $f_{SC}$ in (30) equal to this decision point shows the error region for $\delta_1$. Assume that $\delta_1$ is a real Gaussian random variable with zero mean and variance $\sigma_\delta^2$, hence the probability of correct classification for each supper class is

$$p_S(i|i) = 1 - \int_{x_1}^{x_2} \frac{1}{\sigma_\delta \sqrt{2\pi}} e^{-\frac{\delta_1^2}{2\sigma_\delta^2}} d\delta_1, \tag{31}$$

where $[x_1, x_2]$ is the undesired region for $\delta_1$. In Fig. 4 we illustrate the analytic probability of correct classification for super classes versus $\sigma_\delta^2$ and compare it with the simulation results. In simulation results we estimate our super class feature via (25). For recognizing the modulations types of the received signals the decision is based on (26). All results are generated from 2000 Monte Carlo trials, i.e., 6000 trials for three super classes.

## B. Example 2: Sensitivity for super class with three modulation types from four candidates

If we assume that the number of transmitted signals is $M = 3$, and $Q = 4$, $\alpha_2 = \alpha_1 + \epsilon_1$ and $\alpha_3 = \alpha_1 + \epsilon_2$ where $\epsilon_1, \epsilon_2$ are the difference factors of unknown amplitudes of the received signals. Therefore we can write (29) as

$$f_{SC} = \frac{|\alpha_1|^4 C_{42}(x_1(n)) + |\alpha_1 + \epsilon_1|^4 C_{42}(x_2(n)) + |\alpha_1 + \epsilon_2|^4 C_{43}(x_3(n))}{\left(\frac{|\alpha_1|^2 + |\alpha_1 + \epsilon_1|^2 + |\alpha_1 + \epsilon_2|^2}{3}\right)^2}$$

$$= \frac{C_{42}(x_1(n)) + |1 + \delta_1|^4 C_{42}(x_2(n)) + |1 + \delta_2|^4 C_{43}(x_3(n))}{\left(\frac{1 + |1 + \delta_1|^2 + |1 + \delta_2|^2}{3}\right)^2}, \quad (32)$$

where $\delta_1 = \frac{\epsilon_1}{\alpha_1}$ and $\delta_2 = \frac{\epsilon_2}{\alpha_1}$. By using (32) the undesired region for $\delta_1, \delta_2$ will be found, which if they were placed in that region, $f_{SC}$ will deviate from the correct decision bounds. According to (26), due to use the minimum distance of estimated super class feature from theoretical feature of each super class, to make decision, the decision point between two super classes is the mean of their theoretical features. In (32) if we substitute $f_{SC}$ by the decision point and replace the theoretical cumulant value of modulations in the corresponding supper class in right side of (32), we can find the boundaries of undesired region for each super class. We assume three transmitters with different modulation types can be selected form $\{16 - QAM, BPSK, QPSK, 4 - PAM\}$. Therefore, we have four super classes such as $Sc_1 = \{BPSK, QPSK, 4 - PAM\}$, $Sc_2 = \{16 - QAM, QPSK, 4 - PAM\}$, $Sc_3 = \{BPSK, 16 - QAM, 4 - PAM\}$ and $Sc_4 = \{BPSK, QPSK, 16 - QAM\}$. The decision regions for the above super classes are bounded as follows

$$\delta_{1,1} = -1 \pm 0.01492 \left(4690 + 4690\, \delta_{2,1} + 2345\, \delta_{2,1}^2 \pm 67(-795 + 8460\, \delta_{2,1} - 1590\, \delta_{2,1}^2 - 5820\, \delta_{2,1}^3 - 1455\, \delta_{2,1}^4)^{\frac{1}{2}}\right)^{\frac{1}{2}} \quad (33)$$

$$\delta_{1,2} = -1 \pm 0.0135 \left(4144 + 4144\ \delta_{2,2} + 2072\ \delta_{2,2}^2 \pm 74\big(2100 + 648\ \delta_{2,2} - 10452\ \delta_{2,2}^2 - $$
$$10776\ \delta_{2,2}^3 - 2694\ \delta_{2,2}^4\big)^{\frac{1}{2}}\right)^{\frac{1}{2}} \qquad (34)$$

$$\delta_{1,3} = -1 \pm 0.01492 \left(4690 + 4690\ \delta_{2,3} + 2345\ \delta_{2,3}^2 \pm 67\big(813 + 14892\ \delta_{2,3} + 8058\ \delta_{2,3}^2 + $$
$$612\ \delta_{2,3}^3 + 153\ \delta_{2,3}^4\big)^{\frac{1}{2}}\right)^{\frac{1}{2}} \qquad (35)$$

$$\delta_{1,3} = -1 \pm .00238 \left(161734 + 161734\ \delta_{2,3} + 80867\ \delta_{2,3}^2 \pm 1257\big(-3650 + 48008\ \delta_{2,3} + $$
$$19516\ \delta_{2,3}^2 - 4488\ \delta_{2,3}^3 1122\ \delta_{2,3}^4\ \big)^{\frac{1}{2}}\ \right)^{\frac{1}{2}} \qquad (36)$$

$$\delta_{1,4} = -1 \pm .04347 \left(1288 + 1288\ \delta_{2,4} + 644\ \delta_{2,4}^2 \pm 23\ \big(537 + 4524\ \delta_{2,4} + 1074\ \delta_{2,4}^2 - $$
$$1188\ \delta_{2,4}^3 - 297\ \delta_{2,4}^4\big)^{\frac{1}{2}}\right)^{\frac{1}{2}} \qquad (37)$$

$$\delta_{1,4} = -1 \pm 0.00884 \left(43618 + 43618\ \delta_{2,4} + 21809\ \delta_{2,4}^2 \pm 339\ \big(9298 + 29896\ \delta_{2,4} + $$
$$18596\ \delta_{2,4}^2 + 3648\ \delta_{2,4}^3 + 912 \delta_{2,4}^4\big)^{\frac{1}{2}}\ \right)^{\frac{1}{2}} \qquad (38)$$

where $\delta_{1,k}$, $\delta_{2,k}$ are $\delta_1$, $\delta_2$ for $k$-th super class, respectively. These bounds are shown in Fig. 5. The colored areas are the undesired regions for each super class. $\delta_1$, $\delta_2$ are independent real Gaussian random variable with zero mean and variance of $\sigma_\delta^2$, therefore the probability of correct classification for each supper class is

$$p_S(i|i) = \iint_{\Re} \frac{1}{\sigma_\delta^2\ 2\pi} e^{-\frac{\delta_1^2 + \delta_1^2}{2\sigma_\delta^2}}\ d\delta_1\ d\delta_2 \qquad (39)$$

where $\Re$ is the desired region. In Fig. 6 we illustrate the probability of correct classification of super classes versus $\sigma_\delta^2$ analytically and comparing them with the simulation results. In simulation results we estimate our super class feature via (25). For recognizing the modulations types of the received signals the decisions were made based on (26). All results are according to 2000 Monte Carlo trials, i.e., 8000 trials for four super classes. As it can be seen, analytical result is confirmed by the simulations.

## VI. SENSITIVITY OF PROPOSED CLASSIFIER WITH RESPECT TO THE SYNCHRONIZATION ERROR

According to (2), for rectangular pulse shapes, a synchronization error of $\varepsilon_{T_k}$ $(0 \leq \varepsilon_{T_k} < 1)$ translates, after matched filtering, to an equivalent two-path channel $h_k = [1 - \varepsilon_{T_k} \;\; \varepsilon_{T_k}]$ where $h_k$ is the corresponding channel to $k$-th transmitter [3]. Other pulse shapes also lead to a two-path channel, but the channel coefficients will not be linear in $\varepsilon_{T_k}$ [3]. To show the effect of synchronization error on our proposed classifiers we used simulation results. We assume that $\varepsilon_{T_k}$, $k = 1, 2, \ldots, M$ are random variables and varies from one realization to another independently. They have the uniform distribution in the interval $[0, \vartheta]$, where $\vartheta$ is a positive constant called asynchronous interval.

### A. The effect of synchronization error on SUMC

In this subsection we investigate the effect of synchronization error on the performance of our proposed SUMC based on simulation. Our assumptions in this section are the same as mentioned at the section III for simulations except that we add the effect of synchronization error. In Fig. 7 we illustrated the average probability of correct classification versus asynchronous interval $\vartheta$, with different modulation types such as BPSK, QPSK, $4 - $ PAM and $16 - $ QAM. This figure illustrates the robustness of SUMC with respect to the synchronization error where asynchronous interval $\vartheta$, be less than 0.3.

### B. The effect of synchronization error on MUMC

In this subsection we examine the effect of synchronization error on the performance of the proposed MUMC based on the simulations. These simulation results are shown in Fig. 8. It is illustrated the average probability of correct classification for super class, versus asynchronous interval $\vartheta$, for four super class problem such as $Sc_1 = \{\text{BPSK}, \text{QPSK}, 4 - \text{PAM}\}$, $Sc_2 = \{16 - \text{QAM}, \text{QPSK}, 4 - \text{PAM}\}$, $Sc_3 = \{\text{BPSK}, 16 - \text{QAM}, 4 - \text{PAM}\}$ and $Sc_4 = \{\text{BPSK}, \text{QPSK}, 16 - \text{QAM}\}$ where SNR $= 20$ dB, $N = 6000, 10000$. This figure illustrates the good performance of MUMC with respect to synchronization error where asynchronous interval $\vartheta$, be less than 0.12 that we have only about 12 percents degradations on probability of correct classification.

## VII. CONCLUSION

In this paper we investigated the negative effect of interference on AMC. We have proposed two cumulant based classifier in the presence of interference at two different circumstances. When the received power of one transmitter is larger than the other transmitters, we used single user modulation classification (SUMC) approach to recognize the modulation type of that transmitter's signal and other transmitters are treated as interferences. Alternatively when the received powers of all transmitters are close to each other we proposed multiuser modulation classification (MUMC) method to recognize the modulation type of all of the transmitted signals. We have shown the negative effect of interference on the performance of conventional cumulant based classifier [3], and have illustrated that we can combat this negative effect by using SUMC. In MUMC our proposed classifier has a good performance in the worst case that all of the received signals have the same power. In the case of unequal powers, we assumed the amplitude of the received signals are Gaussian random variable and found the decision boundaries as a function of the variance of these amplitudes. Then, we evaluated the average probability of correct classification for super class versus its normalized variance. It is shown the reasonable performance of the proposed classifier for received signals with different powers. Analytic results are confirmed the computer simulations. It have been shown the robustness of SUMC with respect to the synchronization error where asynchronous interval $\vartheta$, be less than 0.3 and the reasonable performance of MUMC where asynchronous interval $\vartheta$, be less than 0.12.

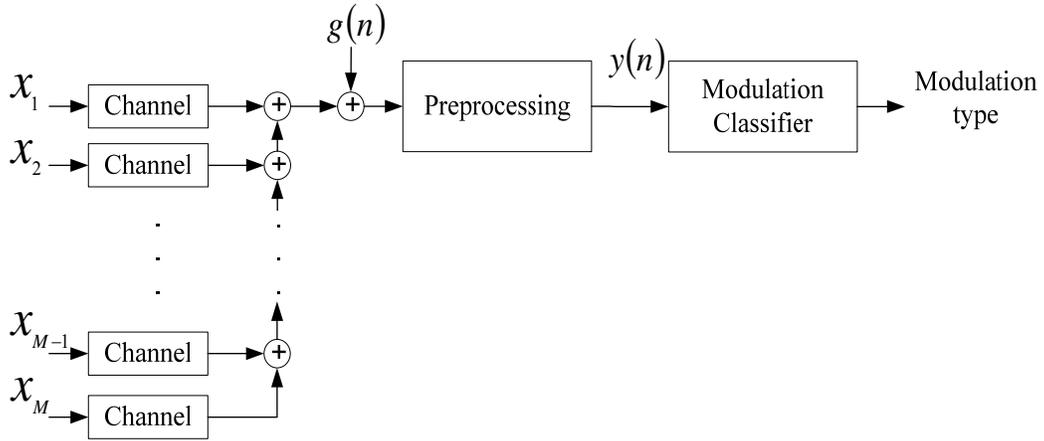

Fig. 1. The block diagram of modulation classification system in the presence of interference.

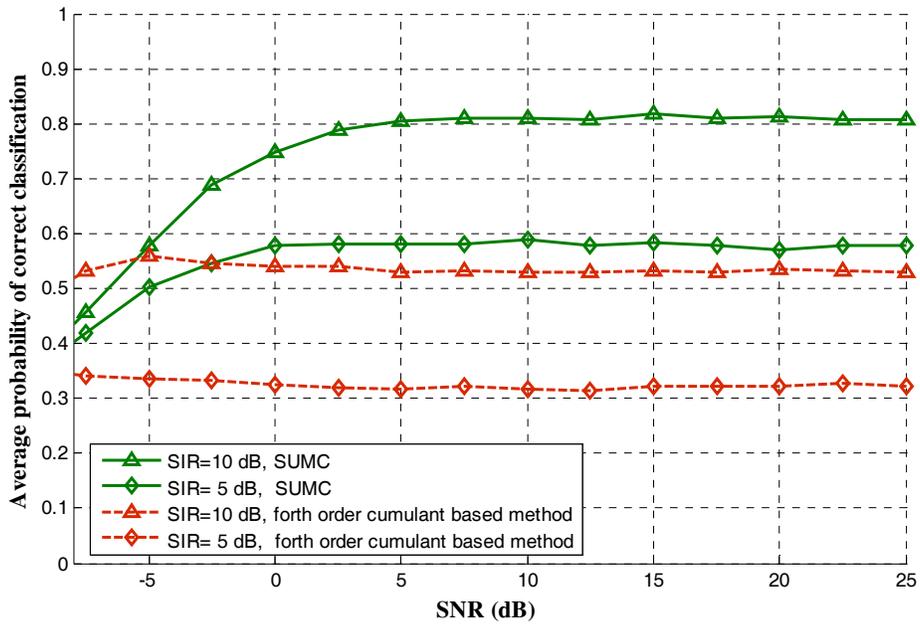

Fig. 2. Average probability of correct classification $p_{cc}$ versus $SNR$ in SUMC manner, whit $SIR = 5, 10\ dB$ for $N = 2000$. Performance of forth order cumulant based and SUMC based classifier is shown by dashed and solid lines respectively.

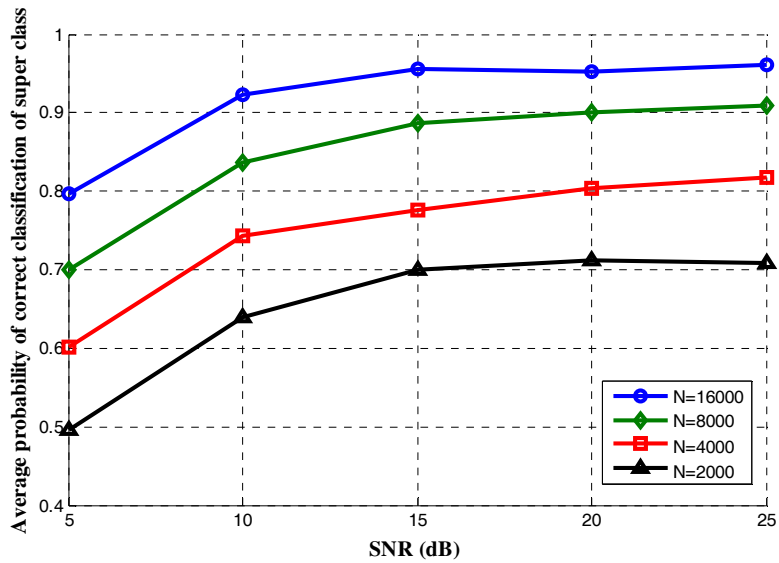

Fig. 3. Average probability of correct classification of super class, $p_{Scc}$ versus $SNR$ in the worst case that the power of all of received signals are equal, with curves parameterized by the number of symbols N for four super class problem such as $\{BPSK, QPSK, 4-PAM\}$, $\{16-QAM, QPSK, 4-PAM\}$, $\{BPSK, 16-QAM, 4-PAM\}$ and $\{BPSK, QPSK, 16-QAM\}$.

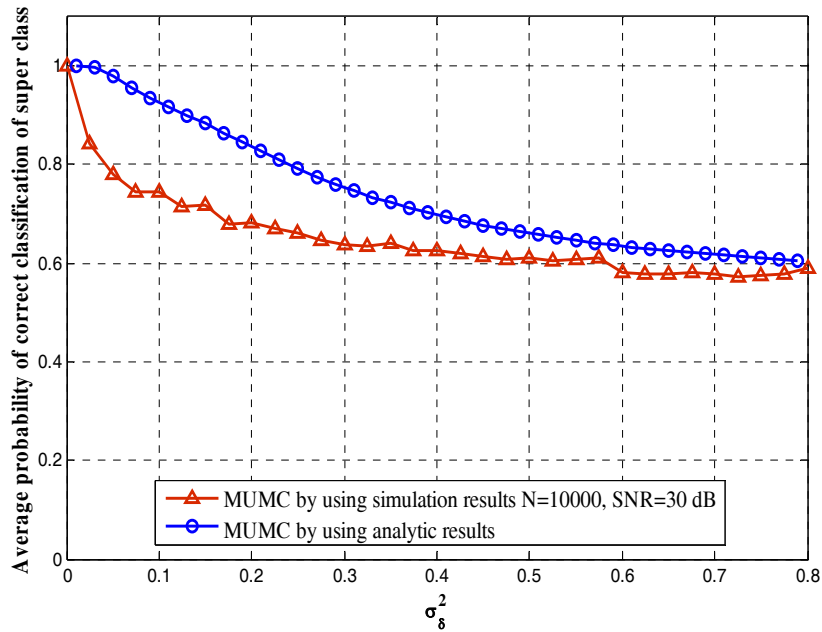

Fig. 4. Average probability of correct classification of super class, $p_{Scc}$ versus the variance of $\delta_1$ of ($\sigma_\delta^2$), with simulation and analytic approaches for three super class such as $Sc_1 = \{BPSK, QPSK\}$, $Sc_2 = \{QPSK, 4-PAM\}$ and $Sc_3 = \{BPSK, 4-PAM\}$.

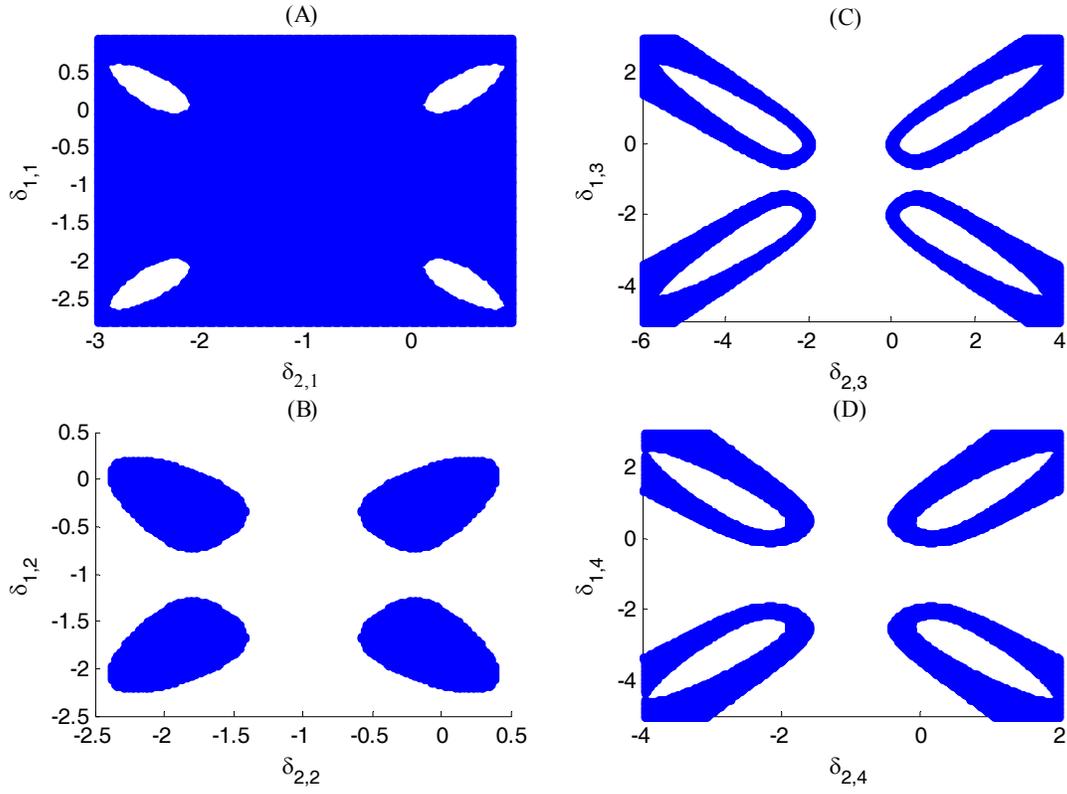

Fig. 5. Decision area for MUMC that the candidate modulations are {BPSK, QPSK, 4 − PAM, 16 − QAM}. Colored region is error region for each supper class. (A) Decision area for $Sc_1$ = {BPSK, QPSK, 4 − PAM}, (B) Decision area for $Sc_2$ = {16 − QAM, QPSK, 4 − PAM}, (C) Decision area for $Sc_3$ = {BPSK, 16 − QAM, 4 − PAM} and (D) Decision area for $Sc_4$ = {BPSK, QPSK, 16 − QAM}.

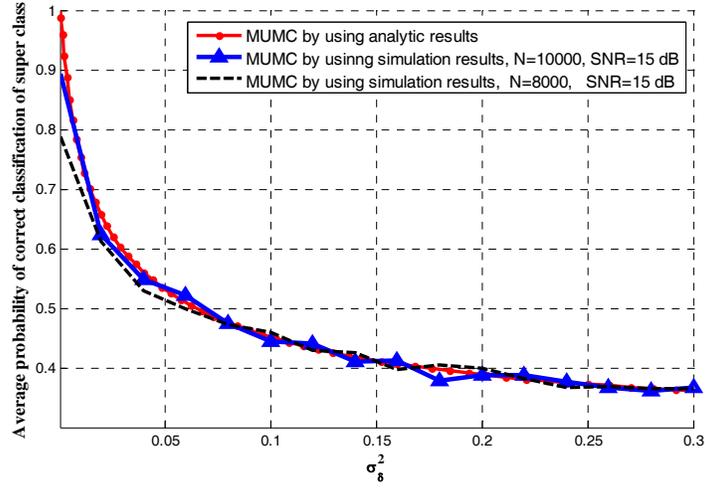

Fig. 6. Average probability of correct classification of super class $p_{Scc}$ versus $\sigma_\delta^2$ with simulation and analytic approaches for four super class problem such as $Sc_1 = \{\text{BPSK}, \text{QPSK}, 4-\text{PAM}\}$, $Sc_2 = \{16-\text{QAM}, \text{QPSK}, 4-\text{PAM}\}$, $Sc_3 = \{\text{BPSK}, 16-\text{QAM}, 4-\text{PAM}\}$ and $Sc_4 = \{\text{BPSK}, \text{QPSK}, 16-\text{QAM}\}$.

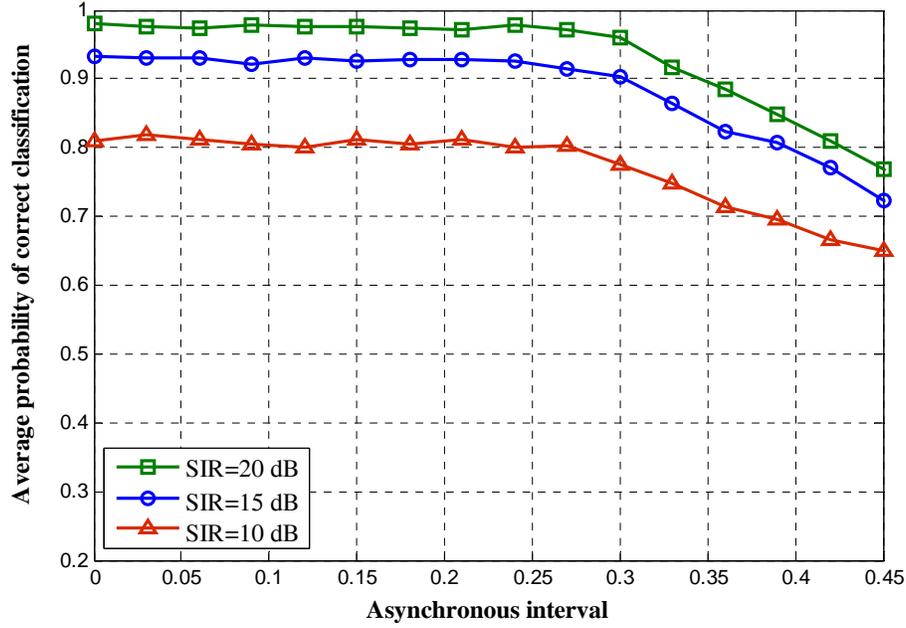

Fig. 7. Average probability of correct classification versus asynchronous interval $\vartheta$, where $\varepsilon_{T_k} \sim U(0, \vartheta)$ for $N = 2000$, $SNR = 15$ dB, $SIR = 10, 15, 20$ dB in SUMC manner by using $\left|\frac{C_{42}(y)}{(C_{63}(y))^{\frac{2}{3}}}\right|$ as feature.

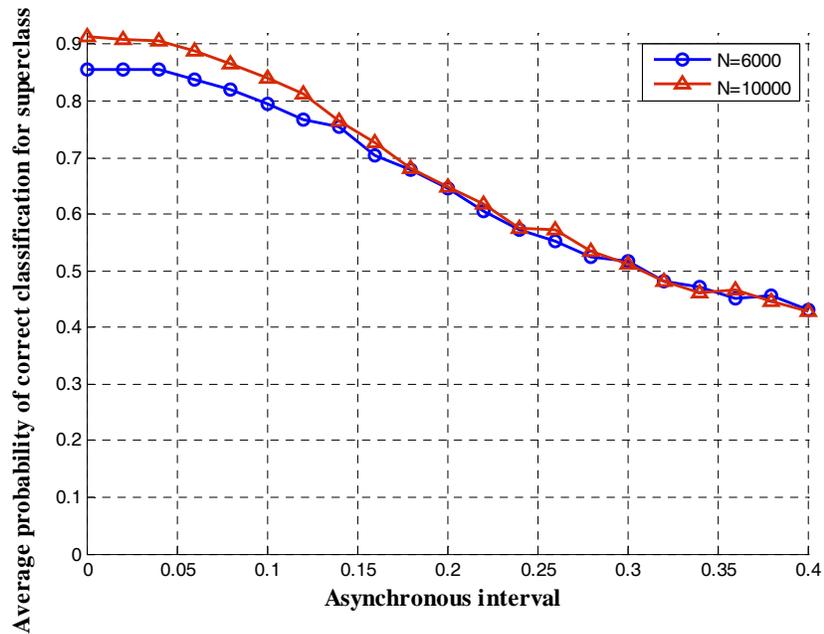

Fig. 8. Average probability of correct classification for super class versus asynchronous interval $\vartheta$, where $\varepsilon_{T_k} \sim U(0, \vartheta)$, SNR = 20 dB four super class problem such as {BPSK, QPSK, 4 − PAM}, {16 − QAM, QPSK, 4 − PAM}, {BPSK, 16 − QAM, 4 − PAM} and {BPSK, QPSK, 16 − QAM}.

TABLE I

THEORETICAL FEATURE UNDER CONSTRAINT OF UNIT VARIANCE.

| Modulation Type  Feature | BPSK | QPSK | 4-PAM | 16-QAM |
|---|---|---|---|---|
| $C_{42}$ | -2 | -1 | -1.36 | -0.68 |
| $C_{63}$ | 16 | 4 | 8.32 | 2.08 |
| $f_c$ | 0.3150 | 0.3969 | 0.3312 | 0.4173 |